\pdfoutput=1 

\documentclass[letterpaper]{article}

\usepackage{microtype}
\usepackage{graphicx}
\usepackage{subfigure}
\usepackage{booktabs} %

\usepackage{framed}
\usepackage{hyperref}
\usepackage{breakurl}
\hypersetup{
           breaklinks=true,   %
           colorlinks=true,   %
           pdfusetitle=true,  %
        }
        
\usepackage{scalefnt}

\usepackage[accepted]{icml2024}

\usepackage{amsmath}
\usepackage{amssymb}
\usepackage{mathtools}
\usepackage{amsthm}

\usepackage[capitalize,noabbrev]{cleveref}

\usepackage{microtype}
\usepackage{inconsolata}
\usepackage{graphicx}
\usepackage{subfigure}
\usepackage{booktabs} %
\usepackage{amssymb}
\usepackage{bm}
\usepackage{bbm}
\usepackage{amsmath}  %
\usepackage{amsthm}
\usepackage{xcolor, color}
\usepackage{textcomp}
\usepackage{multirow}
\usepackage{mathtools}
\usepackage{algorithm}
\usepackage{algpseudocode}
\usepackage{listings,multicol}  %

\usepackage{rotating} 

\algrenewcommand\algorithmicrequire{\textbf{Input:}}
\algrenewcommand\algorithmicensure{\textbf{Output:}}

\usepackage{xspace}
\usepackage{pifont}
\newcommand{\xmark}{\ding{55}}
\usepackage{wrapfig}
\usepackage{tikz}
\usepackage{pifont}
\usepackage{pgfplots}
\usepackage[labelfont=bf,font=small]{caption}
\usepackage{wasysym}

\usetikzlibrary{arrows}
\usetikzlibrary{positioning}
\tikzset{
  treenode/.style = {align=center, inner sep=0pt, text centered,
    font=\sffamily},
  arn_n/.style = {treenode, circle, black, font=\sffamily\bfseries, draw=black,
    fill=white, text width=1.5em},%
  arn_r/.style = {treenode, circle, black, font=\sffamily\bfseries, draw=black,
    fill=white, text width=1.0em},%
  arn_x/.style = {treenode, rectangle, draw=black,
    minimum width=0.5em, minimum height=0.5em}%
}

\definecolor{tabblue}{HTML}{4e79a7}
\definecolor{tabred}{HTML}{e15759}
 \usepackage{url}

\usepackage{bm}
\usepackage{enumitem}

\usepackage{subcaption}
\usepackage{fontawesome}

\usepackage[compact]{titlesec}
\usepackage{etoolbox}
\newcommand{\zerodisplayskips}{%
  \setlength{\abovedisplayskip}{4pt}%
  \setlength{\belowdisplayskip}{4pt}%
  \setlength{\abovedisplayshortskip}{4pt}%
  \setlength{\belowdisplayshortskip}{4pt}}
\appto{\normalsize}{\zerodisplayskips}
\appto{\small}{\zerodisplayskips}
\appto{\footnotesize}{\zerodisplayskips}
\usepackage[subtle, mathdisplays=normal, charwidths=normal, leading=normal]{savetrees}
\algblock{ParFor}{EndParFor}
\algnewcommand\algorithmicparfor{\textbf{parfor}}
\algnewcommand\algorithmicpardo{\textbf{do}}
\algnewcommand\algorithmicendparfor{\textbf{end\ parfor}}
\algrenewtext{ParFor}[1]{\algorithmicparfor\ #1\ \algorithmicpardo}
\algrenewtext{EndParFor}{\algorithmicendparfor}

\definecolor{bred}{RGB}{250, 82, 82}
\definecolor{borange}{RGB}{253, 126, 20}
\definecolor{byellow}{RGB}{250, 176, 5}
\definecolor{bgreen}{RGB}{116, 184, 22}
\definecolor{bblue}{RGB}{250, 176, 5}
\definecolor{bindigo}{RGB}{76, 110, 245}
\definecolor{bcyan}{RGB}{59, 201, 219}
\definecolor{bteal}{RGB}{99, 230, 190}

\usepackage{amsmath,amsfonts,bm}

\def\eqref#1{equation~\ref{#1}}

\def\1{\bm{1}}

\def\rmK{{\mathbf{K}}}

\def\rmO{{\mathbf{O}}}

\def\rmQ{{\mathbf{Q}}}

\def\rmS{{\mathbf{S}}}

\def\rmV{{\mathbf{V}}}

\DeclareMathAlphabet{\mathsfit}{\encodingdefault}{\sfdefault}{m}{sl}
\SetMathAlphabet{\mathsfit}{bold}{\encodingdefault}{\sfdefault}{bx}{n}

\theoremstyle{plain}

\theoremstyle{definition}

\theoremstyle{remark}

\usepackage{xurl}

\usepackage[textsize=tiny]{todonotes}

\icmltitlerunning{Gated Linear Attention Transformers with Hardware-Efficient Training}

\begin{document}

\twocolumn[
\icmltitle{A Study on the Impact of Environmental Liability Insurance on Industrial Carbon Emissions}

\icmlsetsymbol{equal}{*}

\begin{icmlauthorlist}
\icmlauthor{Bo Wu}{yyy,equal}

\end{icmlauthorlist}

\icmlaffiliation{yyy}{Beijing International Studies University}

\icmlcorrespondingauthor{Bo Wu}{\url{2021221257@stu.bisu.edu.cn}}

\icmlkeywords{Machine Learning, ICML}

\vskip 0.3in
]

\printAffiliationsAndNotice{\icmlEqualContribution} %
\vspace{-2mm}
\begin{abstract}
\vspace{-1mm}
In order to explore whether environmental liability insurance has an important impact on industrial emission reduction, this paper selects provincial (city) level panel data from 2010 to 2020 and constructs a two-way fixed effect model to analyze the impact of environmental liability insurance on carbon emissions from both direct and indirect levels. The empirical analysis results show that: at the direct level, the development of environmental liability insurance has the effect of reducing industrial carbon emissions, and its effect is heterogeneous. At the indirect level, the role of environmental liability insurance is weaker in areas with developed financial industry and underdeveloped financial industry. Further heterogeneity analysis shows that in the industrial developed areas, the effect of environmental liability insurance on carbon emissions is more obvious. Based on this, countermeasures and suggestions are put forward from the aspects of expanding the coverage of environmental liability insurance, innovating the development of environmental liability insurance and improving the level of industrialization.

\textbf{key words:  Environmental liability insurance; industrial carbon dioxide emission}
 
\end{abstract}

\section{Introduction}

\subsection{Research Background}

Since 2013, China has consistently ranked at the top globally in terms of carbon emissions across various countries and regions. By 2019, China's carbon emissions accounted for approximately 26\% of the global total, making it the largest emitter, twice that of the second-ranked United States. Among these emissions, carbon dioxide generated from industrial production constitutes the largest proportion of China's total carbon emissions. However, China's industrial competitiveness has not achieved a status commensurate with its carbon emissions. According to the United Nations' Competitive Industrial Performance (CIP) Index, which evaluates a country's actual capacity to create wealth through industry across six dimensions—manufacturing production capacity, manufacturing export capacity, impact on global manufacturing trade, industrial intensity, and export quality—China's CIP ranking rose from 23rd in 2003 to 2nd in 2020, though it still lags behind the top-ranked Germany. Furthermore, after the CIP Index incorporated the dimension of industrial production's environmental impact in 2018, China's adjusted CIP ranking dropped from 3rd to 20th\cite{XCXB202405002}. This indicates that China has become a "carbon emissions giant" before establishing itself as an industrial powerhouse. The high levels of carbon emissions have caused significant global harm, primarily in two ways: (1) Glacier melting due to rising temperatures has led to sea level rise, triggering floods that submerge coastal cities and result in economic losses. (2) The greenhouse effect has caused abnormal temperature increases during rice-growing seasons, affecting grain yields, while also leading to declining water levels that further impact food production. In summary, the issue of carbon emissions cannot be overlooked.

\vspace{-2mm}
\subsection{Research Significance}
\vspace{-2mm}
Industrial carbon emissions primarily stem from two sources. The first is the energy sector. As a major coal-producing country, China not only possesses abundant coal resources but also ranks first globally in coal exports. Coal-fired power generation is economically convenient, yet its low efficiency and high emissions during production make it a primary target for energy sector efforts in energy conservation and emission reduction. The second source is the manufacturing sector. In the production process, factories burn fossil fuels to provide the high temperatures required for chemical reactions, resulting in a portion of carbon dioxide emissions from fossil fuel combustion. Notably, carbon dioxide emitted from cement manufacturing accounts for approximately 5\% of global carbon emissions. Thus, reducing the input of production factors is one approach to lowering industrial carbon emissions. However, if the government mandates low-carbon production through legislation, it poses certain risks. For instance, companies with outdated production equipment and methods, while striving to meet low-carbon and environmental standards, may reduce production inputs, leading to significant decreases in output and substantial profit losses. This could expose them to operational risks. Alternatively, to maintain profits, companies might adopt methods to evade government oversight, creating moral hazards. Moreover, environmental pollution during production could lead to liability for compensation\cite{ZGRZ202110009}. Therefore, enhancing risk protection for companies during production is a crucial method to promote low-carbon production. This has given rise to environmental liability insurance, an essential component of the risk protection system. This insurance transfers the risks of losses due to environmental pollution to insurers, helping companies upgrade equipment and improve production methods while maintaining productivity to the greatest extent possible. However, on the flip side, because this insurance covers compensation for environmental pollution, it may encourage some companies to "lie flat," completely disregarding low-carbon production. To address this, restrictions should be imposed on the insurance premiums. Excessively high premiums may deter companies from purchasing the insurance, forcing them to reduce production under stringent carbon emission limits, which ultimately harms social production and economic development. Conversely, overly low premiums could lead companies to widely adopt this insurance, ignoring carbon emission restrictions and exacerbating environmental pollution. Based on the above analysis, studying the value of carbon insurance in corporate environmental protection holds practical significance.

\vspace{-3mm}
\subsection{Literature Review}

\vspace{-2mm}
\paragraph{Studies on Carbon Emissions} In the context of the "dual carbon" goals (carbon peak and carbon neutrality), carbon peaking, carbon emissions, and their influencing factors have been key research directions for scholars both domestically and internationally. For instance, Liu Minglei (2011) calculated total carbon emissions by multiplying the consumption of various energy types by their corresponding carbon emission coefficients. Yang Yuan (2012), Li Guozhi (2010), and others derived total carbon emissions by selecting the end-use consumption of several primary energy sources. Building on this, Ren Zhijuan (2014) approached the issue from the perspective of regional economic development and resource endowments, highlighting the significant regional disparities in China’s carbon emissions. Zhao Zhe (2022)\cite{CJLC202211005} conducted an in-depth study from the perspective of local fiscal expenditure, finding that larger fiscal expenditures correlate with higher carbon emissions. Moreover, an increase in energy consumption weakens the carbon reduction effects of fiscal expenditure structures. Xie Zhengxuan et al. (2023)\cite{JSNY202306033} discovered that agricultural insurance plays a significant role in reducing agricultural carbon emissions.

\vspace{-2mm}
\paragraph{Studies on Environmental Liability Insurance} Huang Ying (2023) utilized the Smith model to conclude that environmental liability insurance helps enhance corporate environmental awareness, strengthens oversight of companies, and alleviates pressure on governments and nations regarding environmental protection. Dong Hao (2023) further focused on chemical enterprises, finding that environmental liability insurance disperses risks for these companies and reduces financial pressures following pollution incidents. Zhu Jiaxin (2022)\cite{1023014706.nh} targeted ammonia refrigeration companies, analyzing the issue from the perspective of premium amounts and ultimately determining the premium level that maximally mitigates environmental pollution.

\vspace{-2mm}
\paragraph{Commentary.} A review of relevant literature reveals that current research on industrial carbon emissions is relatively comprehensive, with a focus on analyzing how to advance the "dual carbon" goals. However, the scope of studies on specific measures for energy conservation and emission reduction remains narrow. Regarding environmental liability insurance, most research concentrates on pricing levels or its impact on specific companies. In light of this, this study statistically analyzes the historical trends and characteristics of industrial carbon emissions in China. It theoretically examines how environmental liability insurance, as part of a risk protection system, influences overall industrial carbon emissions. Additionally, it empirically tests the effects of environmental liability insurance on industrial carbon emissions, followed by a heterogeneity analysis. This approach enriches the theoretical research on industrial carbon emissions issues.

\vspace{-2mm}

\section{Theoretical Analysis and Research Hypotheses}

\subsection{Theoretical Analysis}
The primary sources of industrial carbon emissions lie in the combustion of fossil fuels and the calcination of raw materials during industrial production. The coverage of environmental liability insurance includes losses incurred by third parties due to pollution of water sources, land, or air caused by accidental incidents during the insured enterprise's operations within the specified scope and region outlined in the insurance contract. The mechanism by which this insurance impacts carbon emissions and environmental pollution operates mainly by providing protection against the adverse consequences of industrial production\cite{FBZX202319011}. This, in turn, alters producers' inputs of fossil fuels and production methods, thereby reducing carbon emissions.

\subsection{Direct Impact Effects}
Environmental liability insurance influences industrial carbon emissions in three main ways: First, it promotes the scale-up of production enterprises. By dispersing operational risks (i.e., losses to third parties caused by accidental pollution of water, land, or air), the insurance enables rapid production recovery, offering producers operational security and enhancing their risk resilience. This encourages producers to expand their production scale. As production scales up, the input of production factors and fossil fuel combustion decrease due to economies of scale. Second, it drives advancements in industrial production technology\cite{1014220591.nh}. The rigorous assessment of accidental incidents under environmental liability insurance compels companies to upgrade production equipment, directly reducing carbon emissions caused by equipment issues without significantly affecting production efficiency. Third, it leverages premium adjustment mechanisms to encourage enterprises to strengthen environmental risk management and improve environmental governance standards. Based on this analysis, this study proposes the following research hypothesis:

\textbf{H1: The development of environmental liability insurance reduces industrial carbon emissions, with heterogeneous effects.}

\subsection{Indirect Market Impact Effects}
As a financial instrument, environmental liability insurance plays a role in energy conservation and emission reduction. Consequently, the level of financial development in a region can either constrain or enhance the effectiveness of this insurance in achieving energy-saving and emission-reduction goals. In regions with advanced financial systems, the coverage rate of environmental liability insurance is higher. Accordingly, the following hypothesis is proposed:

\textbf{H2: The energy-saving and emission-reduction effects of environmental liability insurance are stronger in regions with advanced financial development.}

\section{Research Design}

\begin{table*}[htbp]
	\centering
	\caption{Descriptive Statistics of Variables}
	\label{Descriptive Statistics of Variables}
	\resizebox{\textwidth}{!}{
		\begin{tabular}{l l r r r r}
			\hline
			\textbf{Variable Name} & \textbf{Definition} & \textbf{Mean} & \textbf{Median} & \textbf{Min} & \textbf{Max} \\
			\hline
			Industrial Carbon Emissions (Y) & Calculated using CO\textsubscript{2} conversion coefficient & 42,564 & 32,194.8 & 4,128 & 151,524 \\
			Development Level of Environmental Liability Insurance (Lninsurance) & Log of environmental liability insurance premium income & 23.70 & 23.70236 & 23.14 & 24.19 \\
			Fiscal Support Intensity (FS) & Government expenditure on environmental protection & 143.2 & 115.3478 & 14.89 & 747.4 \\
			Population Size (PS) & Number of permanent residents & 4,585 & 3,894.5 & 563 & 12,624 \\
			Green Area (GA) & Vertical projection area of all vegetation within a certain land area & 102,911 & 75,638.22 & 3,409 & 584,449 \\
			Industrialization Level (LI) & Ratio of industrial added value to regional GDP & 0.332 & 0.3395032 & 0.101 & 0.556 \\
			Urbanization Rate (UR) & Proportion of urban population in total permanent population & 0.584 & 0.5684367 & 0.338 & 0.896 \\
			Proportion of Electricity Consumption (PE) & Share of electricity consumption in total energy consumption & 0.131 & 0.1310803 & 0.0649 & 0.210 \\
			\hline
		\end{tabular}
	}
\end{table*}
\subsection{Data Sources}

This study selects panel data from 32 provinces or regions across China spanning 2010 to 2020. The selected data are sourced from the China Energy Statistical Yearbook, the National Bureau of Statistics, and other relevant sources. Missing data have been supplemented using the linear interpolation method.

\textbf{Table~\ref{Descriptive Statistics of Variables}} presents the descriptive statistical results of the main variables, including their minimum values, maximum values, means, and medians. From the table, it can be observed that there is a significant difference between the minimum and maximum values of industrial carbon emissions, but the mean and median are relatively close, indicating that, overall, the level of industrial development across provinces is similar, though the gap in industrial development levels between certain provinces is pronounced, consistent with the distribution characteristics of industrialization level data. The mean urbanization rate is slightly higher than the median, and the large difference between the maximum and minimum values suggests that, apart from a few provinces with significant wealth disparities, development across most provinces is relatively balanced\cite{CJLC202211005}. The development level of environmental liability insurance, determined by the premium income of this insurance type, shows no significant differences across provinces after logarithmic transformation eliminates the influence of extreme values. The mean fiscal support strength exceeds the median, indicating that a concentrated group of provinces has relatively high fiscal expenditure on environmental protection. Similarly, the mean population size is greater than the median, suggesting that a concentrated group of provinces has a larger number of permanent residents.

\subsection{Research Method}
To investigate the specific impact of the development of environmental liability insurance on industrial carbon emissions, this study uses panel data from various provinces (cities) from 2010 to 2020 and constructs a two-way fixed effects model for regression analysis. The specific expression of the model is as follows:

\[
Y_{it} = \beta_0 + \beta_1 \text{insurance}_{it} + \beta_2 X_{it} + \mu_i + \lambda_t + \varepsilon_{it}
\]

In the above equation:
\begin{itemize}
	\item \( i \) represents each province;
	\item \( t \) represents the year;
	\item \( Y_{it} \) denotes the industrial carbon emissions of province \( i \) in year \( t \);
	\item \( \text{insurance}_{it} \) indicates the development level of environmental liability insurance in province \( i \) in year \( t \);
	\item \( X_{it} \) represents control variables, including fiscal support intensity (FS), population size (PS), industrialization level (LI), urbanization rate (UR), proportion of electricity consumption (PE), and green coverage area (GA);
	\item \( \mu_i \) denotes province fixed effects;
	\item \( \lambda_t \) denotes time fixed effects;
	\item \( \varepsilon_{it} \) represents the random error term.
\end{itemize}

Based on the theoretical analysis in the previous section, it is expected that the coefficient \( \beta_1 \) in this equation is negative, indicating that the development of environmental liability insurance has a negative effect on industrial carbon emissions.

%##################################################################################################
\begin{figure}[t]
	\centering
	\hspace{-1.5mm}
	\includegraphics[width=0.8\linewidth]{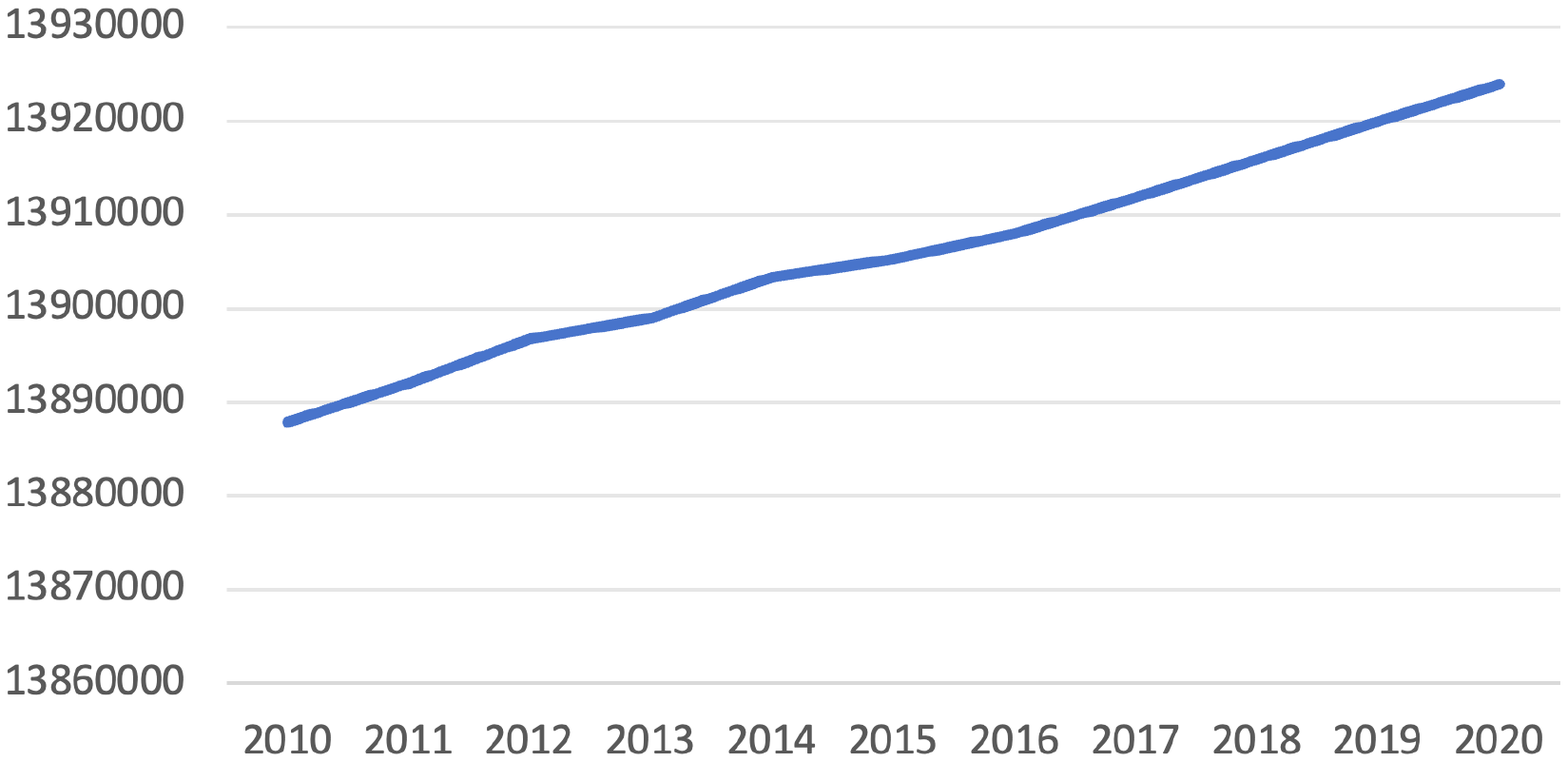}
	\caption{Industrial carbon emissions}
	\label{Industrial}
	\vspace{-.5mm}
\end{figure}
%##################################################################################################
\begin{figure}[t]
	\centering
	\hspace{-1.5mm}
	\includegraphics[width=0.8\linewidth]{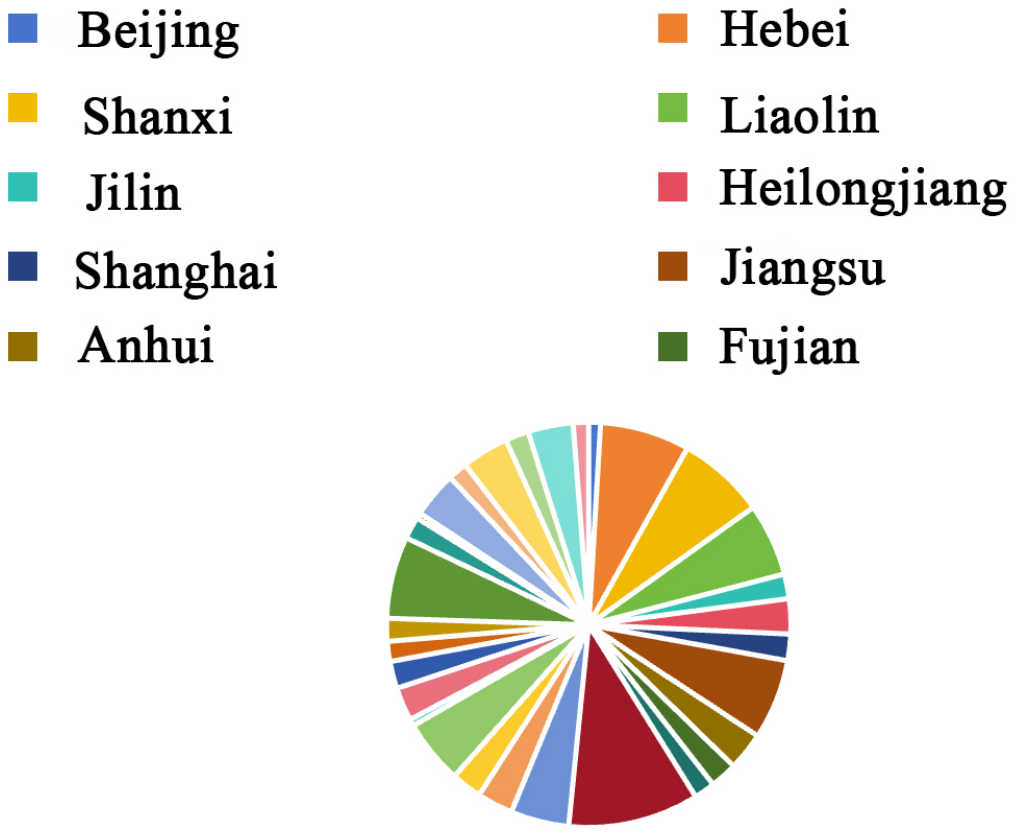}
	\caption{Total industrial carbon emissions by province from 2010 to 2020}
	\label{carbon}
	\vspace{-.5mm}
\end{figure}
%##################################################################################################

%##################################################################################################

 \begin{table}[htbp]
 	\centering
 	\begin{tabular}{lc}
 		\hline
 		\textbf{Energy Type} & \textbf{CO\textsubscript{2} Conversion Factor} \\
 		\hline
 		Coal & 1.9003 \\
 		Tar & 2.8604 \\
 		Crude Oil & 3.0202 \\
 		Gasoline & 2.9251 \\
 		Kerosene & 3.0179 \\
 		Diesel & 3.0959 \\
 		Fuel Oil & 3.1750 \\
 		Liquefied Petroleum Gas & 3.1013 \\
 		Natural Gas & 21.622 \\
 		\hline
 	\end{tabular}
 	\caption{CO\textsubscript{2} Conversion Factors for Different Energy Types}
 	\label{co2conversion}
 \end{table}

%##################################################################################################
%
%
 
%
%
%
%
%
%
%
%
%
%
%
%
%
%
%
%
%
%
%

%
%
%
%
%
%
%
%
%
%
%
%
%
%
%
%
%
%
%
%
%
%
%

%
%
%
%
%
%
%
%
%
%
%
%
%

%
%
%
%
%
%
%
%
%
%
%
%
%
%
%
%
%
%
%
%
%
            
%
%
%
%
%
%
%
%

%
%
%
%
%
%
        
%
%

\subsection{Variable Selection}

\paragraph{Explained Variable Y} : Total Industrial Carbon Dioxide Emissions. Carbon dioxide is currently the primary cause of environmental pollution, including but not limited to air pollution, water pollution caused by ocean acidification, and more. Industrial carbon emissions are mainly reflected in the carbon dioxide gas produced from the consumption of fossil energy during production processes\cite{NKJJ202305008}. According to the classification standards of the China Energy Statistical Yearbook, fossil energy consumption is categorized into nine types: coke, coal, gasoline, kerosene, diesel, crude oil, fuel oil, liquefied petroleum gas, and natural gas. In this study, the measurement of industrial carbon emissions is primarily conducted using the carbon emission coefficient method. Therefore, the total industrial carbon dioxide emissions are calculated by multiplying the consumption of the above nine energy types by their respective carbon dioxide conversion coefficients and then summing the results. The total carbon emissions of province \( i \) in year \( t \) are represented by \( C_{it} \), calculated as follows:

\[
C_{it} = \sum_{k=1}^{9} E_{itk} \times \eta_k
\]

In this equation:
\begin{itemize}
	\item \( C_{it} \) denotes the total carbon emissions of province \( i \) in year \( t \);
	\item \( E_{itk} \) represents the consumption of the \( k \)-th type of fossil energy in province \( i \) in year \( t \);
	\item \( \eta_k \) indicates the carbon dioxide conversion coefficient for the \( k \)-th type of energy, with specific values provided in \textbf{Table~\ref{co2conversion}}.
\end{itemize}

Since the original statistical data for the consumption of various energy types are measured in physical quantities, they must be converted into standard statistical quantities when calculating carbon emissions.

The industrial carbon emissions of various provinces (cities and autonomous regions) from 2010 to 2020 are calculated. As shown in \textbf{Figure~\ref{Industrial}}, from the overall trend of carbon emissions, industrial carbon emissions exhibit an increasing trend, though the rate of increase is relatively small. As shown in \textbf{Figure~\ref{carbon}}, in terms of the distribution of carbon emissions, Hebei, Shanxi, Liaoning, Jiangsu, Shandong, Henan, Guangdong, and Shaanxi provinces all exceed the average level, while Hainan and Qinghai provinces have relatively low emissions.
\paragraph{Core Explanatory Variable}Level of Environmental Liability Insurance Development. The development level of environmental liability insurance is primarily reflected in the premium income of this insurance type. To mitigate the impact of extreme data on empirical results, this data is logarithmically processed in this study.  

\paragraph{Other Control Variables}Numerous factors influence industrial carbon emissions. Drawing on the research findings of *The Carbon Emission Reduction Effects of Local Fiscal Expenditure in China*, this study selects six factors as control variables: fiscal support strength, population size, urbanization rate, industrialization level, proportion of electricity consumption, and green coverage area. Specifically: 
 
\textbf{(1) Fiscal Support Strength} is represented by fiscal expenditure on environmental protection, which falls under non-economic fiscal expenditure. Increasing the proportion of non-economic fiscal expenditure is conducive to suppressing carbon emissions. With the proposal of the "dual carbon" goals, the proportion of national fiscal expenditure on environmental protection relative to total fiscal expenditure has been steadily increasing, thus exerting a certain negative impact on industrial carbon emissions; the expected coefficient is negative.  

\textbf{(2) Population Size} refers to the number of permanent residents. The larger the urban population, the greater the scale of industrial production activities, directly affecting carbon emissions; the expected coefficient is positive.  

\textbf{(3) Urbanization Rate} refers to the proportion of the urban population to the total population. The population structure of a region is an important factor influencing carbon emissions; the expected coefficient is negative.  

\textbf{(4) Industrialization Level}  is measured by the proportion of industrial added value to the regional gross domestic product. Missing data (e.g., Yunnan Province in 2020) have been supplemented using linear interpolation. Since industrial production is the primary source of industrial carbon emissions, this proportion has a direct impact on industrial carbon emissions; the expected coefficient is positive.  

\textbf{(5) Proportion of Electricity Consumption}: According to the \textbf{2023 Carbon Emission Rankings of China’s Top 100 Listed Companies}, the power, cement, and steel industries account for a significant share of carbon emissions, with average carbon efficiencies of 0.07 million yuan/ton, 0.08 million yuan/ton, and 0.32 million yuan/ton, respectively—all below the list’s average—indicating substantial pressure for carbon reduction. Among these, the power sector is the most prominent, emitting approximately 379 million tons of carbon dioxide\cite{1019233209.nh}. Therefore, whether the power generation industry prioritizes energy saving and emission reduction is of utmost importance; the expected coefficient is positive. 
 
\textbf{(6) Green Coverage Area} refers to the vertical projection area of all plants within a specific land area. Since green plants can absorb carbon dioxide from the air, green coverage area has a negative impact on carbon emissions; the expected coefficient is negative.

\section{Empirical Study}
%##################################################################################################
\begin{figure}[t]
	\centering
	\hspace{-1.5mm}
	\includegraphics[width=0.8\linewidth]{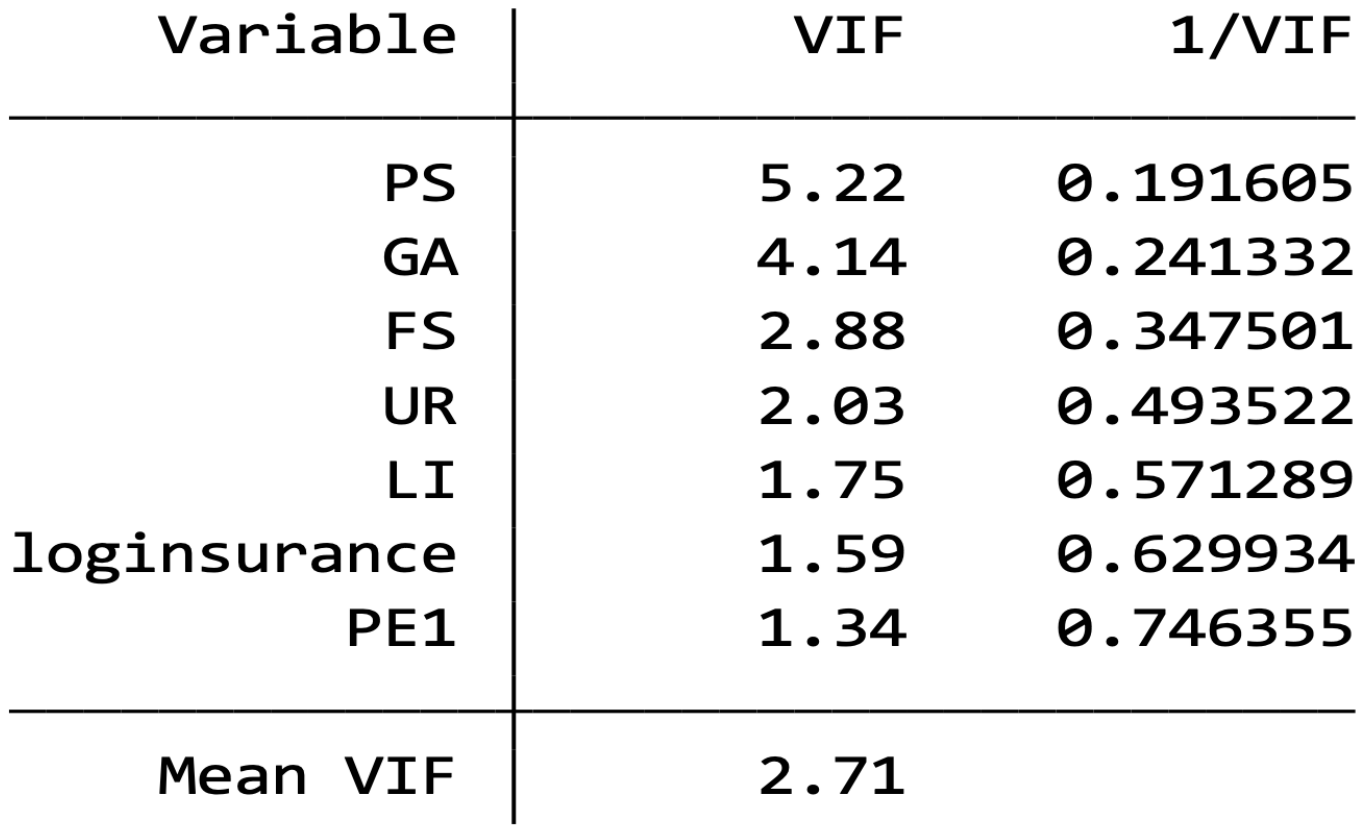}
	\caption{\textbf{Multicollinearity test}}
	\label{Multicollinearity}
	\vspace{-.5mm}
\end{figure}

%##################################################################################################
\subsection{Multicollinearity Test}

To ensure that the selected variables meet the standards for empirical regression, a multicollinearity test was conducted on the variables. As shown in \textbf{Figure~\ref{Multicollinearity}}, the maximum VIF value is 5.22, with a mean of 2.71, both of which are less than 10. This indicates that there is no multicollinearity among the selected variables, allowing for regression analysis.

\begin{table*}[htbp]
	\centering
	\caption{Regression Results of Industrial Carbon Emissions}

    \label{RegressionResults}
	\resizebox{\textwidth}{!}{
			\begin{tabular}{lccc}
			\toprule
			Variable Name & Industrial Carbon Emissions (1) & Industrial Carbon Emissions (2) & Industrial Carbon Emissions (3) \\
			\midrule
			Development Level of Environmental Liability Insurance & 8,526 & 14,089.218 & -5,854.696* \\
			Lninsurance & (7,800) & (7,725.940) & (-1.96) \\
			Fiscal Support Intensity &  & 37.073 & -7.329 \\
			FS &  & (22.683) & (-0.96) \\
			Population Size &  & 3.053** & 0.890 \\
			PS &  & (1.054) & (0.39) \\
			Level of Industrialization &  & 48,870.830* & 33,562.547*** \\
			LI &  & (19,923.527) & (3.62) \\
			Level of Urbanization &  & -2,338.808 & 40,250.228** \\
			UR &  & (14,796.100) & (2.14) \\
			Proportion of Electricity Consumption &  & -211,617.786*** & 134,122.028*** \\
			PE &  & (48,829.751) & (3.61) \\
			Green Coverage Area &  & 0.074** & 0.056** \\
			GA &  & (0.026) & (2.10) \\
			Constant & -159,466 & -305,311.091 & 88,701.191 \\
			& (184,821) & (181,077.522) & (1.24) \\
			\midrule
			Observations & 330 & 330 & 330 \\
			Adjusted R-squared &  &  & 0.972 \\
			Province FE &  &  & YES \\
			Year FE &  &  & YES \\
			\bottomrule
		\end{tabular}
	}
	\vspace{0.5em}
	\caption*{\footnotesize Standard errors in parentheses. *** p$<$0.01, ** p$<$0.05, * p$<$0.1}
\end{table*}
\subsection{Benchmark Model}

Through the Hausman test and multicollinearity test, the issue of multicollinearity was ruled out, and a fixed effects model was selected for empirical analysis.

Based on the model
\[
Y_{it} = \beta_0 + \beta_1 \text{insurance}_{it} + \beta_2 X_{it} + \mu_i + \lambda_t + \varepsilon_{it}
\]

Before fixing the time effects and provincial effects, there was no linear relationship between carbon emissions and the core explanatory variable (the development level of environmental liability insurance). This may be influenced by differences across provinces.

After fixing the two variables—province and year—the regression results are shown in \textbf{Table~\ref{RegressionResults}}. The core explanatory variable, the development level of environmental liability insurance, passed the significance test at the 10\% level, with a negative regression coefficient. This indicates that environmental liability insurance has a negative impact on industrial carbon emissions, confirming its role in reducing energy consumption and emissions in industrial carbon emissions.

\paragraph{The industrialization level is significant at the 1\% level,}with a positive regression coefficient, indicating a positive correlation between industrialization level and industrial carbon emissions. Since China's industrial revolution started relatively late and industrial production efficiency remains low, maintaining high industrial productivity for development requires increased inputs of raw materials such as fossil fuels. Therefore, the higher the level of industrialization, the greater the industrial carbon emissions.

\paragraph{The proportion of electricity consumption is significant at the 1\% level,}with a positive regression coefficient, suggesting that the proportion of electricity consumption is directly proportional to industrial carbon emissions. As of 2023, thermal power generation accounts for as much as 66.3\% of total power generation in China, and thermal power produces far more carbon compounds than wind, hydropower, or other generation methods. Thus, the higher the proportion of electricity consumption, the greater the industrial carbon emissions.
\vspace{-2mm}
\paragraph{The urbanization rate is significant at the 5\% level,}with a positive regression coefficient, indicating a positive correlation between urbanization rate and industrial carbon emissions. Unlike Western countries where industrialization drives urbanization, China, despite not undergoing an industrial revolution, has a large population and abundant cheap labor as a demographic dividend. This has led to urbanization driving industrial development. On one hand, a higher urbanization rate corresponds to a higher level of industrial development, resulting in greater industrial carbon emissions. On the other hand, the influx of large populations into cities has spurred the real estate industry and infrastructure construction. During the processes of building and demolition, the extensive use of machinery, fuel, and electricity significantly increases industrial carbon emissions.
\vspace{-2mm}
\paragraph{The green coverage area is significant at the 5\% level.}However, its regression coefficient is positive, suggesting a positive correlation between green coverage area and industrial carbon emissions. This may be because green areas are located far from industrial enterprises, making them unable to effectively absorb the carbon dioxide released from industrial production.
$\parallel$In summary, these regression results largely confirm \textbf{ Hypothesis 1 (H1)}: the development of environmental liability insurance contributes to reducing industrial carbon emissions, with its effects exhibiting heterogeneity.

\subsection{Heterogeneity Analysis}

\begin{table*}[htbp]
	\centering
	\caption{Regression Results for Financially Developed vs Underdeveloped Regions}
	
	\label{RegressionFinancially}
	\resizebox{\textwidth}{!}{
		\begin{tabular}{lcc}
			\toprule
			Variable Name & Industrial Carbon Emissions \\
			& in Financially Developed Regions & in Financially Underdeveloped Regions \\
			\midrule
			Development Level of Environmental Liability Insurance & -4,760.444 & -8,609.792** \\
			& (-1.14) & (-2.44) \\
			Fiscal Support Intensity & 15.461 & 7.839 \\
			& (1.64) & (0.75) \\
			Population Size & -1.866 & 10.894** \\
			& (-0.76) & (2.49) \\
			Level of Industrialization & -48,155.243*** & 57,233.931*** \\
			& (-3.19) & (5.48) \\
			Urbanization Rate & 114,640.643*** & -119,652.290*** \\
			& (4.82) & (-2.81) \\
			Proportion of Electricity Consumption & -53,549.811 & 342,886.161*** \\
			& (-0.89) & (6.44) \\
			Green Coverage Area & 0.024 & 0.025 \\
			& (0.85) & (0.58) \\
			Constant & 60,867.069 & 151,999.234* \\
			& (0.61) & (1.76) \\
			\midrule
			Observations & 121 & 209 \\
			Adjusted R-squared & 0.978 & 0.979 \\
			Province FE & YES & YES \\
			Year FE & YES & YES \\
			\bottomrule
		\end{tabular}
	}
	\vspace{0.5em}
	\caption*{\footnotesize Standard errors in parentheses. *** p$<$0.01, ** p$<$0.05, * p$<$0.1}
\end{table*}
\paragraph{Financially Developed vs Underdeveloped Regions}
Since environmental liability insurance serves as a financial instrument in promoting energy conservation and emission reduction, the level of financial development in a region can, to some extent, either constrain or enhance its effectiveness in achieving these goals. To investigate whether the energy-saving and emission-reduction effects of environmental liability insurance differ between provinces with advanced and underdeveloped financial sectors, this study conducts a subsample analysis of various provinces. The level of financial development in a province or region is assessed based on the ratio of total deposits and loans to GDP. Using the average value of this indicator as a threshold, the 32 provinces are divided into financially developed regions and financially underdeveloped regions. The regression results for these two groups are presented in \textbf{Table~\ref{RegressionFinancially}}.

Contrary to \textbf{Hypothesis 2 (H2)}—which posits that environmental liability insurance has a stronger effect on energy conservation and emission reduction in financially developed regions—the empirical results suggest the opposite. In financially developed regions, the effect of environmental liability insurance is weaker compared to financially underdeveloped regions. In financially underdeveloped regions, environmental liability insurance passes the significance test at the 5\% level and exerts a negative impact on industrial carbon emissions\cite{FBZX202319011}. The reasons for this can be attributed to three main factors:

Moral Hazard: Financially developed regions typically have higher levels of economic development and per capita consumption, along with stronger production incentives. As a result, the effectiveness of environmental liability insurance in these areas is limited.

Information Asymmetry: Based on past cases involving environmental liability insurance, disputes often arise over the definition of "accidental" events (the insurance covers sudden and accidental pollution incidents, while intentional or malicious pollution is excluded). Whether a pollution incident is accidental or intentional is known only to the company’s internal personnel, while insurance companies struggle to investigate such matters effectively.

Sophistication of Enterprises: Enterprises in financially developed regions tend to have extensive experience in the financial sector, a deeper understanding of financial instruments, and greater proficiency in their use. This undoubtedly increases the difficulty for insurance companies to conduct investigations.

\begin{table*}[htbp]
	\centering
	\caption{Regression Results for Industrially Developed vs Underdeveloped Regions}
	
	\label{RegressionIndustrially}
	\resizebox{\textwidth}{!}{
		\begin{tabular}{lcc}
			\toprule
			Variable Name & Industrial Carbon Emissions \\
			& in Industrially Developed Regions & in Industrially Underdeveloped Regions \\
			\midrule
			Development Level of Environmental Liability Insurance & -13,191.071*** & -1,813.598 \\
			& (-2.82) & (-0.49) \\
			Fiscal Support Intensity & 14.821 & -22.615* \\
			& (1.18) & (-1.98) \\
			Population Size & -18.931*** & 7.530* \\
			& (-4.34) & (1.81) \\
			Level of Industrialization & 16,425.146 & 31,033.729*** \\
			& (0.85) & (2.97) \\
			Urbanization Rate & -235,271.762*** & 29,788.694 \\
			& (-3.23) & (1.37) \\
			Proportion of Electricity Consumption & 193,238.635** & 123,401.395*** \\
			& (2.43) & (3.08) \\
			Green Coverage Area & 0.176*** & -0.005 \\
			& (3.64) & (-0.15) \\
			Constant & 543,310.922*** & -1,259.704 \\
			& (4.34) & (-0.01) \\
			\midrule
			Observations & 165 & 165 \\
			Adjusted R-squared & 0.974 & 0.960 \\
			Province FE & YES & YES \\
			Year FE & YES & YES \\
			\bottomrule
		\end{tabular}
	}
	\vspace{0.5em}
	\caption*{\footnotesize Standard errors in parentheses. *** p$<$0.01, ** p$<$0.05, * p$<$0.1}
\end{table*} 
\paragraph{Different Levels of Industrialization Across Provinces}
Given that the impact of industrial carbon emissions varies with different levels of industrialization, this study conducts a subsample analysis of various provinces. The level of industrial development in a province or region is determined by the proportion of industrial added value to total output value. Using the average value of this indicator as a threshold, the 32 provinces are divided into industrially developed regions and industrially underdeveloped regions. The regression results for these two groups are presented in \textbf{Table~\ref{RegressionIndustrially}}.

The results indicate that in industrially developed regions, environmental liability insurance passes the significance test at the 1\% level and exerts a negative impact on carbon emissions. However, in industrially underdeveloped regions, there is no linear relationship between carbon emissions and the core explanatory variable (the development level of environmental liability insurance). This is because production equipment and methods in industrially underdeveloped regions are relatively outdated, and environmental liability insurance alone cannot drive advancements in production technology to reduce carbon emissions.

\begin{table*}[htbp]
	\centering
	\caption{Robustness Test Results}
	
	\label{RegressionRobustness}
	\resizebox{\textwidth}{!}{
		\begin{tabular}{lccc}
			\toprule
			Variable Name & Carbon Emission Intensity (1) & Carbon Emission Intensity (2) & Carbon Emission Intensity (3) \\
			\midrule
			Development Level of Environmental Liability Insurance & -1.232*** & 0.222 & -37.399** \\
			Lninsurance & (0.401) & (0.405) & (-2.55) \\
			Fiscal Support Intensity &  & 0.005*** & -7.308 \\
			FS &  & (0.001) & (-0.97) \\
			Population Size &  & -0.001*** & 0.909 \\
			PS &  & (0.000) & (0.40) \\
			Level of Industrialization &  & 6.660*** & 33,217.985*** \\
			LI &  & (1.045) & (3.60) \\
			Urbanization Rate &  & -7.071*** & 40,752.339** \\
			UR &  & (0.776) & (2.17) \\
			Proportion of Electricity Consumption &  & -6.836** & 132,550.794*** \\
			PE &  & (2.562) & (3.58) \\
			Green Coverage Area &  & 0.000*** & 0.056** \\
			GA &  & (0.000) & (2.11) \\
			Constant & 31.28*** & 0.971 & -43,025.563** \\
			& (9.496) & (9.502) & (-2.52) \\
			\midrule
			Observations & 330 & 330 & 330 \\
			Adjusted R-squared &  &  & 0.972 \\
			Province FE &  &  & YES \\
			Year FE &  &  & YES \\
			\bottomrule
		\end{tabular}
	}
	\vspace{0.5em}
	\caption*{\footnotesize Standard errors in parentheses. *** p$<$0.01, ** p$<$0.05, * p$<$0.1}
\end{table*}
\subsection{Robustness Test}
Changing the dimension of the dependent variable. The previous section calculated carbon emissions based on the carbon emission coefficient method. To avoid errors caused by a single core dependent variable, such as the impact of economic development on carbon emissions, this study uses industrial carbon emission intensity instead of industrial carbon emission volume. The formula for industrial carbon emission intensity is:

\[
\text{Carbon emission intensity} = \frac{\text{Carbon emissions}}{\text{GDP}}
\]

Using this indicator, a regression analysis was conducted again. The regression results are shown in \textbf{Table~\ref{RegressionRobustness}}. The results indicate that the development level of environmental liability insurance passed the significance test at the 5\% level, with the regression coefficient remaining negative. This confirms the reliability of the baseline regression results, demonstrating that environmental liability insurance has a negative impact on industrial carbon emission intensity.

\vspace{-2mm}
\section{Conclusion}
\vspace{-2mm}
Since China proposed the "dual carbon" goals, environmental liability insurance has emerged as a new tool for reducing industrial carbon emissions. Empirical studies have found that: at a direct level, the development of environmental liability insurance can reduce industrial carbon emissions, though its effects are heterogeneous\cite{1018822194.nh}. At an indirect level, in regions with advanced financial sectors, the role of environmental liability insurance is weaker compared to regions where the financial sector is less developed. Further heterogeneity analysis reveals that in industrially advanced regions, the suppressive effect of environmental liability insurance on carbon emissions is more pronounced.
\subsection{Policy Recommendations}
\paragraph{Expand the Coverage of Carbon Insurance}
Carbon insurance not only mitigates business operational risks but also contributes to environmental protection. Therefore, its adoption should be further promoted. On the government side, there should be policy support and a legislative framework. Governments should formulate policies to encourage the development of carbon insurance, such as offering tax incentives or subsidies, and establish a clear legal and regulatory framework to ensure the transparency and reliability of carbon insurance products\cite{1023662922.nh}. In terms of outreach, public and corporate awareness of carbon insurance should be raised through education and promotional campaigns. Additionally, the importance of carbon risk management should be emphasized within the business community, encouraging companies to adopt carbon insurance as a risk mitigation tool.
\paragraph{Innovate and Develop Diverse Carbon Insurance Products}
Insurance companies should actively develop a variety of carbon insurance products to meet the diverse needs of clients, thereby promoting research and innovation within the industry to better quantify and manage carbon risks. At the same time, standardized systems for collecting and reporting carbon emissions data should be established to enable more accurate risk assessments.
\paragraph{Enhance Industrial Modernization}
Efforts should be made to promote technological innovation by integrating science and technology with industrial production and environmental liability insurance\cite{STJJ202403001}. This can involve developing low-carbon production technologies, using advanced technologies to replace or reduce fossil fuel inputs, and elevating the level of industrial modernization. These measures would improve industrial production efficiency and reduce industrial carbon emissions.

\bibliography{main}
\bibliographystyle{util/icml2024}
\newpage
\appendix

\onecolumn

\end{document}